# DPRAODV: A DYANAMIC LEARNING SYSTEM AGAINST BLACKHOLE ATTACK IN AODV BASED MANET

**Payal N. Raj, Prashant B. Swadas**
**[1] Computer Engineering Department, SVMIT**
**Bharuch, Gujarat, India**
*payalnraj@gmail.com*

**[2] Computer Engineering Department, B.V.M.**
**Anand, Gujarat, India**
*prashantswadas @yahoo.com*

## Abstract

Security is an essential requirement in mobile ad hoc networks to provide protected communication between mobile nodes. Due to unique characteristics of MANETS, it creates a number of consequential challenges to its security design. To overcome the challenges, there is a need to build a multifence security solution that achieves both broad protection and desirable network performance. MANETs are vulnerable to various attacks, blackhole, is one of the possible attacks. Black hole is a type of routing attack where a malicious node advertise itself as having the shortest path to all nodes in the environment by sending fake route reply. By doing this, the malicious node can deprive the traffic from the source node. It can be used as a denial-of-service attack where it can drop the packets later. In this paper, we proposed a DPRAODV (Detection, Prevention and Reactive AODV) to prevent security threats of blackhole by notifying other nodes in the network of the incident. The simulation results in ns2 (ver-2.33) demonstrate that our protocol not only prevents blackhole attack but consequently improves the overall performance of (normal) AODV in presence of black hole attack.

## 1. Introduction

Mobile ad hoc network (MANET) is one of the recent active fields and has received spectacular consideration because of their self-configuration and self-maintenance. Early research assumed a friendly and cooperative environment of wireless network. As a result they focused on problems such as wireless channel access and multihop routing. But security has become a primary concern to provide protected communication between mobile nodes in a hostile environment. Although mobile ad hoc networks have several advantages over wired networks, on the other side they pose a number of non-trivial challenges to the security design as they are more vulnerable than wired networks [1]. These challenges include open network architecture, shared wireless medium, demanding resource constraints, and, highly dynamic network topology. In this paper, we have considered a fundamental security problem in MANET to protect its basic functionality to deliver data bits from one node to another. Nodes help each other in conveying information to and fro and thereby creating a virtual set of connections between each other. Routing protocols play an imperative role in the creation and maintenance of these connections. In contrast to wired networks, each node in an ad-hoc networks acts like a router and forwards packets to other peer nodes. The wireless channel is accessible to both legitimate network users and malicious attackers. As a result, there is a blurry boundary separating the inside network from the outside world.

Many different types of routing protocols have been developed for ad hoc networks and have been classified into two main categories by Royer and Toh (1999) as *Proactive* (periodic) protocols and *Reactive* (on-demand) protocols. In a proactive routing protocol, nodes periodically exchange routing information with other nodes in an attempt to have each node always know a current route to all destinations [2]. In a reactive protocol, on the other hand, nodes exchange routing information only when needed, with a node attempting to discover a route to some destination only when it has a packet to send to that destination [3]. In addition, some ad hoc network routing protocols are hybrids of periodic and on-demand mechanisms.

Wireless ad hoc networks are vulnerable to various attacks. These include passive eavesdropping, active interfering, impersonation, and denial-of-service. A single solution cannot resolve all the different types of attacks in ad hoc networks. In this paper, we have designed a novel method to detect blackhole attack: DPRAODV, which isolates that malicious node from the network. We have complemented the reactive system on every node on the network. This agent stores the Destination sequence number of incoming route reply packets (RREPs) in the routing table and calculates the threshold value to evaluate the dynamic training data in every time interval as in [4]. Our solution makes the participating nodes realize that, one of their neighbors is malicious; the node thereafter is not allowed to participate in packet forwarding operation.

In Section 2 of this paper, we summarize the basic operation of AODV (Ad hoc On-Demand distance Vector Routing) protocol on which we base our work. In Section 3, we discuss related work. In Section 4, we describe the effect of blackhole attack in AODV. Section 5 presents the design of our protocol; DPRAODV that protects against blackhole attack. Section 6 discusses the performance evaluation based on simulation experiments. Finally, Section 7 presents conclusion and future work

## 2. Theoretical background of AODV

AODV is a reactive routing protocol; that do not lie on active paths neither maintain any routing information nor participate in any periodic routing table exchanges. Further, the nodes do not have to discover and maintain a route to another node until the two needs to communicate, unless former node is offering its services as an intermediate forwarding station to maintain connectivity between other nodes [3]. AODV has borrowed the concept of destination sequence number from DSDV [5], to maintain the most recent routing information between nodes.

Whenever a source node needs to communicate with another node for which it has no routing information, Route Discovery process is initiated by broadcasting a Route Request (RREQ) packet to its neighbors. Each neighboring node either responds the RREQ by sending a Route Reply (RREP) back to the source node or rebroadcasts the RREQ to its own neighbors after increasing the hop_count field. If a node cannot respond by RREP, it keeps track of the routing information in order to implement the reverse path setup or forward path setup [6].

The destination sequence number specifies the freshness of a route to the destination before it can be accepted by the source node. Eventually, a RREQ will arrive to node that possesses a fresh route to the destination. If the intermediate node has a route entry for the desired destination, it determines whether the route is fresh by comparing the destination sequence number in its route table entry with the destination sequence number in the RREQ received. The intermediate node can use its recorded route to respond to the RREQ by a RREP packet, only if, the RREQ's sequence number for the destination is greater than the recorded by the intermediate node. Instead, the intermediate node rebroadcasts the RREQ packet. If a node receives more than one RREPs, it updates its routing information and propagates the RREP only if RREP contains either a greater destination sequence number than the previous RREP, or same destination sequence number with a smaller hop count. It restrains all other RREPs it receives. The source node starts the data transmission as soon as it receives the first RREP, and then later updates its routing information of better route to the destination node. Each route table entry contains the following information:

- Destination node
- Next hop
- number of hops
- Destination sequence number
- Active neighbors for the route
- Expiration timer for the route table entry

The route discovery process is reinitiated to establish a new route to the destination node, if the source node moves in an active session. As the link is broken and node receives a notification, and Route Error (RERR) control packet is being sent to all the nodes that uses this broken link for further communication. And then, the source node restarts the discovery process.

As the routing protocols typically assume that all nodes are cooperative in the coordination process, malicious attackers can easily disrupt network operations by violating protocol specification. This paper discusses about blackhole attack and provides routing security in AODV by purging the threat of blackhole attacks

## 3. Related works in securing AODV

There are basically two approaches to secure MANET: (1) Securing Ad hoc Routing and (2) Intrusion Detection [7].

### 3.1 Secure Routing

The Secure Efficient Ad hoc Distance vector routing protocol (SEAD) [8] employs the use of hash chains to authenticate hop counts and sequence numbers in DSDV. Another secure routing protocol, Ariadne[9] assumes the existence of a shared secret key between two nodes based on DSR (reactive) routing protocol. The Authenticated Routing for Ad hoc networks (ARAN) is a standalone protocol that uses cryptographic public-key certificates in order to achieve the security goals [10]. Security-Aware Ad hoc Routing (SAR) uses security attributes such as trust values and relationships [11].

The computation overhead involved in the above mentioned protocols is awful and often suffer from scalability problems. As a preventive measure, the packets are carefully signed, but an attacker can simply drop the packet passing through it, therefore, secure routing cannot resist such internal attacks. So our solution provides a reactive scheme that triggers an action to protect the network from future attacks launched by this malicious node.

## 3.2 Intrusion Detection System

Zhang and Lee [12] present an intrusion detection technique for wireless ad hoc networks that uses cooperative statistical anomaly detection techniques. The use of anomaly based detection techniques results in too many number of false positives. Stamouli proposes architecture for Real-Time Intrusion Detection for Ad hoc Networks (RIDAN) [7]. The detection process relies on a state-based misuse detection system. Therefore, each node requires extra processing power and sensing capabilities.

In [13], the method requires the intermediate node to send Route Confirmation Request (CREQ) to next hop towards the destination. This operation can increase the routing overhead resulting in performance degradation. In [14], source node verifies the authenticity of node that initiates RREP by finding more than one route to the destination, so that it can recognize the safe route to destination. This method can cause the routing delay, since a node has to wait for RREP packet to arrive from more than two nodes. In [4], the feature used is dest_seq_no, which reflects the trend of updating the threshold and hence reflecting the adaptively change in network environment.

Therefore, a method that can prevent the attack without increasing routing overhead and delay is required. All the above mentioned approaches except [4], use static value for threshold. To resolve the problem, threshold value should be reflecting current network environment by updating its value. And also, our solution ensures that a node once detected as malicious cannot participate in forwarding and sending of a data packet in the network.

## 4. Description of Blackhole attack

MANETs are vulnerable to various attacks. General attack types are the threats against Physical, MAC, and network layer which are the most important layers that function for the routing mechanism of the ad hoc network. Attacks in the network layer have generally two purposes: not forwarding the packets or adding and changing some parameters of routing messages; such as sequence number and hop count. A basic attack that an adversary can execute is to stop forwarding the data packets. As a result, when the adversary is selected as a route, it denies the communication to take place. In blackhole attack, the malicious node waits for the neighbors to initiate a RREQ packet. As the node receives the RREQ packet, it will immediately send a false RREP packet with a modified higher sequence number. So, that the source node assumes that node is having the fresh route towards the destination. The source node ignores the RREP packet received from other nodes and begins to send the data packets over malicious node. A malicious node takes all the routes towards itself. It does not allow forwarding any packet anywhere. This attack is called a blackhole as it swallows all objects; data packets [15].

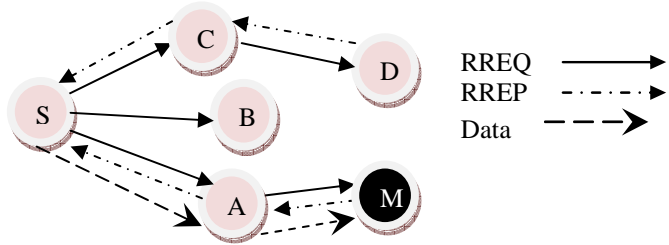

Fig. 1  Blackhole attacks in MANETs

In figure 1, source node S wants to send data packets to a destination node D in the network. Node M is a malicious node which acts as a blackhole. The attacker replies with false reply RREP having higher modified sequence number. So, data communication initiates from S towards M instead of D.

## 5. DPRAODV: Solution against blackhole attack

In normal AODV, the node that receives the RREP packet first checks the value of sequence number in its routing table. The RREP packet is accepted if it has RREP_seq_no higher than the one in routing table. Our solution does an addition check to find whether the RREP_seq_no is higher than the threshold value. The threshold value is dynamically updated as in [4] in every time interval. As the value of RREP_seq_no is found to be higher than the threshold value, the node is suspected to be malicious and it adds the node to the black list. As the node detected an anomaly, it sends a new control packet, ALARM to its neighbors. The ALARM packet has the black list node as a parameter so that, the neighboring nodes know that RREP packet from the node is to be discarded. Further, if any node receives the RREP packet, it looks over the list, if the reply is from the blacklisted node; no processing is done for the same. It simply ignores the node and does not receive reply from that node again. So, in this way, the malicious node is isolated from the network by the ALARM packet. The continuous replies from the malicious node are blocked, which results in less Routing overhead. Moreover, unlike AODV, if the node is found to be malicious, the routing table for that node is not updated, nor the packet is forwarded to another node.

The threshold value is dynamically updated using the data collected in the time interval. If the initial training data were used, then the system could not adapt the changing environment. The threshold value is the average of the difference of dest_seq_no in each time slot between the sequence number in the routing table and the RREP packet. The time interval to update the threshold value is as soon as a newer node receives a RREP packet. As a new node receives a RREP for the first time, it gets the updated value of the threshold. So our design not only detects the blackhole attack, but tries to prevent it further, by updating threshold which reflects the real changing environment. Other nodes are also updated about the malicious act by an ALARM packet, and they react to it by isolating the malicious node from network.

# 6. Evaluation of DPRAODV

## 6.1 Simulation Environment

For simulation, we have used ns2 (v-2.33) network simulator [16]. Mobility scenarios are generated by using a Random waypoint model by varying 10 to 70 nodes moving in a terrain area of 800m x 800m. Each node independently repeats this behavior and mobility is varied by making each node stationary for a period of pause time. The simulation parameters are summarized in Table 1.

Table 1: Simulation Parameters

| Parameter | Value |
|---|---|
| Simulator | Ns-2(ver.2.33) |
| Simulation time | 1000 s |
| Number of nodes | 70 |
| Routing Protocol | AODV |
| Traffic Model | CBR |
| Pause time | 2 (s) |
| Maximum mobility | 60 m/s |
| No. of sources | 5 |
| Terrain area | 800m x 800m |
| Transmission Range | 250m |
| No. of malicious node | 1 |

A new Routing Agent is added in ns-2 to include the blackhole attack. In order to implement blackhole attack, the malicious node generates a random number between 15 and 200, adds the number to the sequence number in RREQ and then generates the sequence number in RRREP. In our simulation, the communication is started between source node to the destination node in presence of the malicious node. The node number of source node, destination node and malicious node are 2, 7 and 0 respectively.

## 6.2 Simulation Evaluation Methodology

The simulation is done to analyze the performance of the network's various parameters. The metrics used to evaluate the performance are given below:

- Packet Delivery Ratio: The ratio of the data delivered to the destination to the data sent out by the source.
- Average End-to-end delay: The difference in the time it takes for a sent packet to reach the destination. It includes all the delays, in the source and each intermediate host, caused by the routing discovery, queuing at the interface queue etc.
- Normalized routing overhead: This is the ratio of routing-related transmissions (RREQ, RREP, RERR etc) to data transmissions in a simulation. A transmission is one node either sending or forwarding a packet. Either way, the routing load per unit data successfully delivered to the destination.

## 6.2 Simulation Analysis and Results

Various network contexts are considered to measure the performance of a protocol. These contexts are created by varying the following parameters in the simulation.

- Network size: variation in the number of mobile nodes.
- Traffic load: variation in the number of sources
- Mobility: variation in the maximum speed

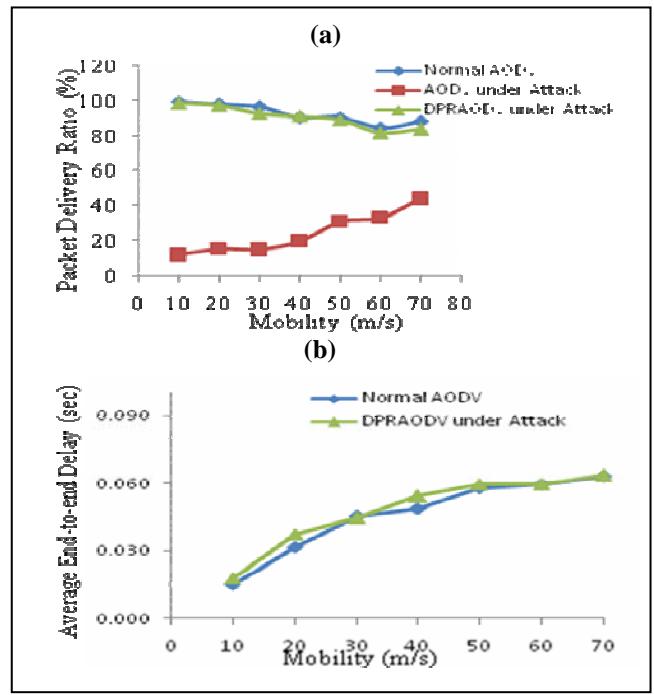

Fig. 2 Impact of Mobility on the performance

Figure 2a and 2b conclude the simulation based on the effect of mobility on the DPRAODV compared to normal AODV. The PDR stays within acceptable limits almost 4-5% lower than it should normally be with minimum overhead.

than AODV under attack with Average-End-to-end delay almost same as normal AODV.

In Figure 3c, it is observed that there is slight increase in Normalized Routing Overhead, which is quite negligible. In AODV under attack, the delay will be less and routing overhead will be quite high compared to normal AODV, so our comparison is between normal AODV and DPRAODV.

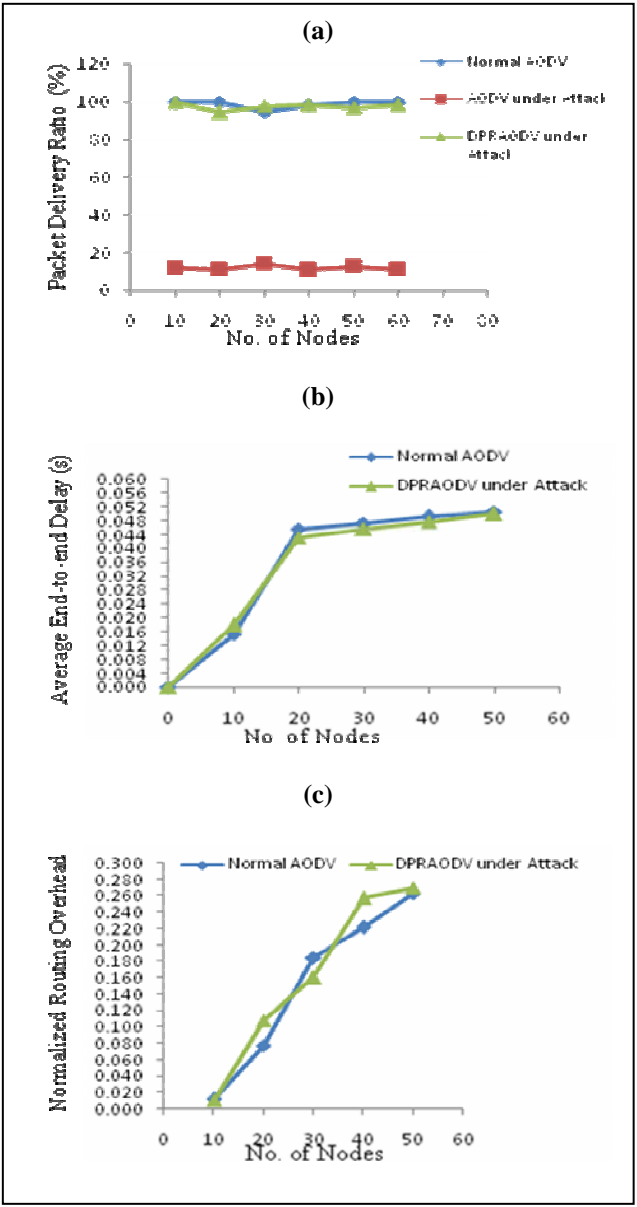

Fig. 3 Impact of Network Size on the performance

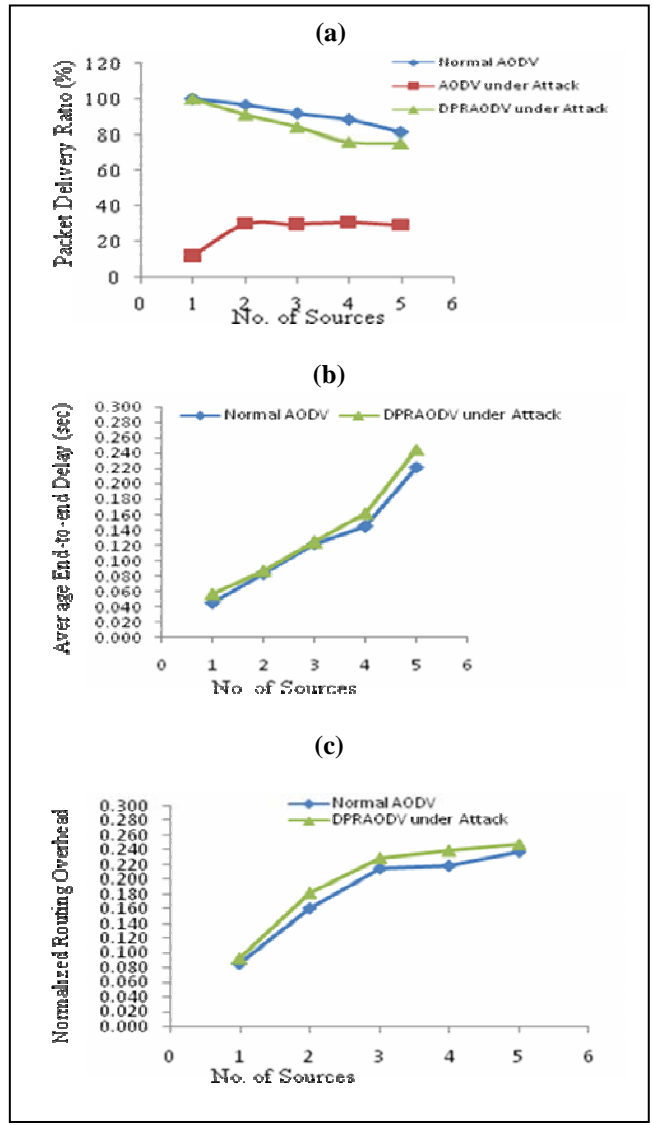

Fig. 4 Impact of Traffic Load on the performance

All the above three contexts are simulated and tested to see the effect of network size on Packet Delivery Ratio( PDR), Average End-to end delay and Normalized Routing Overhead.

From figure 3a and b, we analyze that, under blackhole attack, the PDR of DPRAODV is improved by 80-85%

From the figure 4, it is clear that as the traffic load increases, the PDR of DPRAODV increases by approximately 60% than AODV under attack. As our solution generates ALARM packet, there is slight increase in Normalized Routing Overhead with almost same Delay as normal AODV.

# 7. Conclusions

In DPRAODV, we have used a very simple and effective way of providing security in AODV against blackhole attack. As from the graphs illustrated in results we can easily infer that the performance of the normal AODV drops under the presence of blackhole attack. Our prevention scheme detects the malicious nodes and isolates it from the active data forwarding and routing and reacts by sending ALARM packet to its neighbors. Our solution: DPRAODV increases PDR with minimum increase in Average-End-to-end Delay and normalized Routing Overhead.

### Acknowledgments

This work is sponsored by the Institute of Science and Technology for Advanced Studies and Research (ISTAR), Vallabh Vidyanagar, Gujarat, India.